\documentclass{article}

\usepackage{arxiv}

\usepackage[utf8]{inputenc} 
\usepackage[T1]{fontenc}    
\usepackage{url}            
\usepackage{booktabs}       
\usepackage{amsfonts}       
\usepackage{nicefrac}       
\usepackage{microtype}      
\usepackage[autostyle=true]{csquotes} 

\usepackage{bm}
\usepackage{bbm}
\usepackage{graphics}
\usepackage{hyperref}
\usepackage{dsfont}
\usepackage{enumitem}
\usepackage{amsmath}
\usepackage{amsfonts}
\usepackage{amssymb}
\usepackage{mathtools}
\usepackage{subcaption}
\usepackage[table]{xcolor}
\usepackage{tcolorbox}
\usepackage{pdfpages}
\usepackage[standard]{ntheorem}
\usepackage{microtype}
\usepackage[ruled]{algorithm2e}
\usepackage{array}
\RequirePackage{scrextend}
\RequirePackage{tabularx}

\usepackage[natbibapa]{apacite}
\usepackage{natbib}
\changefontsizes[14]{11}
\newcommand{\pr}{\mathbb{P}}

\newcommand{\one}{\mathds{1}}

\newcommand{\setN}{\mathcal{N}}

\newcommand{\given}{\, |\,}
\newcommand{\Sehat}{\widehat{\Se}}
\newcommand{\Sphat}{\widehat{\Sp}}

\newcommand{\Sebm}{\boldsymbol{\Se}}
\newcommand{\Spbm}{\boldsymbol{\Sp}}

\newcommand{\bb}{\bm{b}}

\DeclareMathOperator{\AUC}{AUC}

\DeclareMathOperator{\se}{se}

\DeclareMathOperator{\Se}{se}
\DeclareMathOperator{\Sp}{sp}

\hypersetup{
    colorlinks,
    breaklinks=true,
    urlcolor={red!50!black},
    linkcolor={red!50!black},
    citecolor={blue!50!black}
}

\title{Statistical Inference for Diagnostic Test Accuracy Studies with Multiple Comparisons}
\author{
  Max Westphal\thanks{The two authors contributed equally and are listed in alphabetical order.} \thanks{Corresponding author: Max Westphal, \href{mailto:mwestphal@uni-bremen.de}{max.westphal@mevis.fraunhofer.de}, \url{https://orcid.org/0000-0002-8488-758X}}\\
  Fraunhofer MEVIS \\
  Institute for Digital Medicine\\ 
  Bremen, Germany \\
   \And
 Antonia~Zapf\footnotemark[1] \\
  Department of Medical Biometry and Epidemiology\\
  University Medical Center Hamburg-Eppendorf\\
  Hamburg, Germany
}

\begin{document}

\maketitle


\begin{abstract}

Diagnostic accuracy studies assess sensitivity and specificity of a new index test in relation to an established comparator or the reference standard. 
The development and selection of the index test is usually assumed to be conducted prior to the accuracy study. In practice, this is often violated, for instance if the choice of the (apparently) best biomarker, model or cutpoint is based on the same data that is used later for validation purposes. 
In this work, we investigate several multiple comparison procedures which provide family-wise error rate control for the emerging multiple testing problem. Due to the nature of the co-primary hypothesis problem, conventional approaches for multiplicity adjustment are too conservative for the specific problem and thus need to be adapted.
In an extensive simulation study, five multiple comparison procedures are compared with regards to statistical error rates in least-favorable and realistic scenarios. This covers parametric and nonparamtric methods and one Bayesian approach. All methods have been implemented in the new open-source R package \texttt{cases} which allows to reproduce all simulation results. 
Based on our numerical results, we conclude that the parametric approaches (maxT, Bonferroni) are easy to apply but can have inflated type I error rates for small sample sizes. The two investigated Bootstrap procedures, in particular the so-called pairs Bootstrap, allow for a family-wise error rate control in finite samples and in addition have a competitive statistical power.

\end{abstract}


\section{Introduction}
The aim of diagnostic and prognostic studies is generally to differentiate between two groups, e.g. the diseased from the non-diseased or those with good outcome from those with poor outcome. This differentiation can be based on a single clinical parameter or a combination of parameters. The methods and results presented here are independent of whether the goal is diagnosis or prognosis and whether a single parameter or a combination is evaluated. Furthermore, this work also covers the scenario that several investigated diagnostic procedures are defined by different (machine-learned) prediction models which are based on the same input data, e.g. medical images. For better readability, the term diagnosis, diagnostic study and diagnostic test is used in the following, but prognosis, prognosis study and prognostic marker are meant at the same time.

In order to assess the accuracy of a new diagnostic test, the so-called index test, the result is compared with the true condition, which is assessed using the gold standard or reference standard. The recommended co-primary endpoints are sensitivity as the proportion correctly diagnosed as diseased and specificity as the proportion correctly diagnosed as non-diseased. If a standard test exists, the index test should be compared with it, preferably in the within-subject design (all tests in all individuals). However, if there is no established comparator, the aim is to demonstrate a pre-specified minimum sensitivity and specificity, the so-called minimally acceptable criteria (MAC) \citep{chmp_guideline_2010, korevaar_targeted_2019}.

Since the study is only considered a success if both hypotheses regarding sensitivity and specificity can be rejected, no correction is required for the multiplicity of the two endpoints \citep{chmp_guideline_2010}. However, multiplicity problems may occur elsewhere in diagnostic studies, especially if the aim is to evaluate different markers in parallel and select the best diagnostic test or to determine the optimal cut-off value for a single marker. Usually, a test or cut-off value is selected in a first step data-driven and then validated in a new study. However, it is known that data-driven selection may lead to an overestimation of the diagnostic accuracy and that the results of the selection study often cannot be replicated in the validation study \citep{leeflang_systematic_2008}. The same phenomenon, sometimes referred to as optimization bias or selection-induced bias, is well known to also exist in bioinformatics and predictive modelling \citep{boulesteix_optimal_2009, jelizarow_over-optimism_2010}.

Therefore, \citet{westphal_evaluation_2020} have proposed a framework for the evaluation of multiple prediction models in which not the (apparently) single-best model is chosen and subsequently validated, but rather a set of promising models from which the best is selected in the evaluation study. \citet{westphal_multiple_2022} have transferred the framework to diagnostic accuracy studies with co-primary endpoints sensitivity and specificity. In the same way, the framework can also be used to determine the optimal cut-off value for a single biomarker after choosing a range of promising cut points in a prior development study. Regardless of the context, this framework offers increased flexibility and and can increase statistical power, i.e. the probability to correctly demonstrate a sufficiently high diagnostic accuracy for one of the selected tests.

However, it is important to correct for multiplicity in the validation study to avoid inflation of the type-one error.
The correction approaches can be divided into single-step and stepwise procedures. In contrast to stepwise procedures, single-step procedures have the advantage that associated simultaneous confidence intervals can be constructed \citep{strassburger_compatible_2008}. This is relevant in diagnostic accuracy studies as confidence intervals are an important indicator of estimation uncertainty.

The simplest single-step correction is the Bonferroni adjustment, in which the adjusted significance level is equal to the global significance level, say $\alpha=0.05$, divided by the number of tests. This approach leads to conservative results when the diagnostic test results are positively correlated. This is for instance the case if the same biomarker or risk score is dichotomized at slightly different cut points.
More elaborate multiple comparison procedures considered in this work aim to exploit the correlation structure and thereby increase statistical power. Existing approaches include the maxT-approach \citep{hothorn_simultaneous_2008} utilized by \citet{westphal_evaluation_2020}. This method is based on a multivariate normal approximation  which leads to less conservative decisions in light of positive correlations. 
Nonparametric methods for diagnostic test evaluation were developed by \citet{konietschke_rank-based_2012} and \citet{zapf_wild_2015} transferred this approach to early diagnostic studies with the area under the receiver operating characteristic (ROC) curve, the so-called AUC as primary outcome measure. \citet{rudolph_adjusting_2017} in turn transferred the approaches from the AUC to sensitivity and specificity as co-primary endpoints.

The main contribution of this work is the adaptation of several multiple comparison procedures to diagnostic accuracy studies with multiple index tests with sensitivity and specificity as co-primary endpoints. The results of an extensive simulation study are presented to allow a systematic, neutral and reproducible comparison of relevant statistical properties. Finally, we introduce a new open-source R package \texttt{cases}\footnote{\url{https://github.com/maxwestphal/cases}, last accessed \today.} which provides convenient implementations for all investigated methods.

This work is structured as follows. 
Section \ref{sec:example} presents a motivational real-world data example. 
Different multiple comparison procedures and the design of the simulation study are introduced in Section \ref{sec:methods}.
Section \ref{sec:results} then presents the results of the simulation study and the example data set. 
A discussion of the results and the conclusion can be found in Section \ref{sec:discussion}.

\section{Motivating example: Breast cancer diagnosis}\label{sec:example}

For the motivating example, we utilize the "Breast Cancer Wisconsin (Diagnostic) Data Set"\footnote{\url{https://archive.ics.uci.edu/ml/datasets/breast+cancer+wisconsin+(diagnostic)} (last assessed \today)}. In total, the dataset consists of 569 observations of digitized image of a fine needle aspirate of a breast masses. Hereby, $n_1 = 212$ observations are breast cancer cases and $n_0 = 357$ are controls (fibrocystic breast masses). Thirty real-valued features that were computed from the image data which describe characteristics of the cell nuclei present in the image. A detailed description of the dataset and feature extraction methodology is provided by \citet{street1993, wolberg1994, wolberg1995, mangasarian1995}. Two (independent) example scenarios are described below and analysed in Section \ref{sec:results_rwd}. For both scenarios, we assume that a high sensitivity is deemed more important compared to a high specificity because false negatives have more serious consequences than false positives.

\subsection{Scenario A: Biomarker assessment}

In this first example scenario, our goal is the assessment of the diagnostic accuracy of multiple biomarker candidates in a single study. In this example, three of the thirty available features, have been selected before the study based on prior data and knowledge. The selected biomarker candidates are the most extreme area, compactness and concavity of any of the ceil nuclei.
To define a diagnostic test based on a continuous biomarker, a threshold needs to be specified in addition to allow categorization into diseased and healthy subjects. However, the optimal threshold can usually only be specified with uncertainty before the study. In our example, we specify five threshold candidates for each biomarker, roughly corresponding to the 30\%, 40\%, 50\%, 60\% and 70\% quantiles of each biomarker distribution.
In this scenario, our framework will be used to investigate the diagnostic accuracy of the resulting 15 combinations of markers (3) and thresholds (5 per marker) simultaneously while adjusting for the resulting multiplicity.

\subsection{Scenario B: Risk model evaluation}

In modern medical applications, individual predictive features are commonly combined to risk prediction models. In this second example scenario, we evaluate several candidate models simultaneously. We split the dataset into 379 and 190 observations for model learning and evaluation, respectively, to emulate a prospective evaluation study after model development. In the development phase, five variants of the elastic net algorithm are employed, corresponding to five different values of the penalty mixing hyperparameter $\alpha \in \{0, 0.25, 0.5, 0.75, 1\}$ \citep{zou2005, friedman2010}. For each of the five variants, the optimal penalty strength $\lambda$ is found via 10-fold-cross-validation. 
In the subsequent evaluation phase, the five optimized models ought to be assessed simultaneously. Because each model provides a predicted probability of cancer, the question arises which probability threshold should be used in practice to make classifications. As a high sensitivity is targeted, the thresholds 10\%, 20\%, 30\%, 40\%, 50\% will be evaluated on the remaining 'prospective' evaluation data. In effect, in this scenario, our framework will be used to investigate the diagnostic accuracy of the resulting 25 combinations of risk models (5) and probability thresholds (5 per model) simultaneously while adjusting for the resulting multiplicity.

\section{Methods}\label{sec:methods}
In the following, the statistical model and the hypotheses are presented (\ref{sec:model}). Various approaches which allow for a multiplicity adjusted analysis are then described (\ref{sec:tests} and \ref{sec:regions}). Finally, the design of the simulation study is outlined (\ref{sec:numerical}).

\subsection{Statistical model and hypotheses}\label{sec:model}
A sample of $n$ individuals is assumed, divided into $n_0$ non-diseased and $n_1$ diseased (with $n=n_0+n_1$). We consider the within-subject design, i.e. for each person $i=1,\ldots,n$, the reference standard and $m$ index tests are applied. The comparison of each test $T_j$, $j=1,\ldots,m$, and reference standard $D$ results in a true sensitivity $\Se_j=P(T_j=1|D=1)$ and specificity $\Sp_j=P(T_j=0|D=0)$.
Hereby, $T_j \in \{0,1\}$ indicates test-positive (1) and test-negative (0) and $D\in \{0,1\}$ indicates diseased (1) and non-diseased (0) individuals, respectively. As stated in the introduction, the different index tests $T_j$ may be based on a single clinical parameter or biomarker which is dichotomized based on different cut points. On the other hand, each $T_j$ may also be the result of a different complex prediction model based on high-dimensional input data (e.g. medical images). A mixture of simple and complex models is also possible. \\
We consider the two alternative hypotheses that true sensitivity and specificity are above a pre-specified minimum sensitivity and specificity, which are denoted by $\Se_0$ and $\Sp_0$ \citep{korevaar_targeted_2019}. This leads to the individual null hypotheses
\begin{align}\label{eq:null}
    H_{0,j}^{\se}: \se_j\leq \se_0, \quad H_{0,j}^{\Sp}: \Sp_j\leq \Sp_0
\end{align}
for each test $T_j$.
These two hypotheses are connected according to the intersection-union principle to the combined null hypothesis
\begin{align}\label{eq:null_combined}
    H_{0,j}:\ H_{0,j}^{\se}\ \cup\ H_{0,j}^{\Sp}
\end{align}
per index test $T_j$, $j=1,\ldots,m$. 
The global null hypothesis for our multiple testing problem then reads as
\begin{align}\label{eq:null_global}
    H_0:\ \bigcap_{j=1}^m H_{0,j} = \bigcap_{j=1}^m \left\{ H_{0,j}^{\se}\ \cup\ H_{0,j}^{\Sp} \right\}
\end{align}

Usual (maximum-likelihood) estimates of sensitivity ($\widehat{se}_j$) and specificity ($\widehat{sp}_j$) are calculated as
\begin{align}\label{eq:null_estimates}
    \widehat{se}_j=\frac{n_{j,11}}{n_1}\quad \mbox{and} \quad \widehat{sp}_j=\frac{n_{j,00}}{n_0}
\end{align}
with $n_{j,11}$ as number of diseased individuals with a positive test result and $n_{j,00}$ as number of non-diseased individuals with a negative test result, each for test $j$. In the \texttt{cases} package, the estimates \eqref{eq:null_estimates} are actually slightly shrunk towards $0.5$ by default to prevent singular (co)variance estimates \cite{westphal_multiple_2022}. The decision to retain or reject any null hypothesis can be made by means of a multiple comparison procedure as discussed in the next two sections.

\subsection{Multiple comparison procedures}\label{sec:tests}

As a starting point for the data analysis we define usual Wald test statistics for the $j$-th index test by
\begin{align}
    Z^{\Se}_j = \frac{\Sehat_j-\se_0}{\sqrt{\Sehat_j(1-\Sehat_j)/n_1}} \quad \mbox{and} \quad 
    Z^{\Sp}_j = \frac{\Sphat_j-\Sp_0}{\sqrt{\Sphat_j(1-\Sphat_j)/n_0}}.
\end{align}
A level $\alpha$ test for the combined hypothesis $H_j$ (without adjustment for multiplicity) can easily be obtained. For this, both individual test statistics need to surpass an appropriate critical value $c_\alpha$ \citep{chmp_guideline_2010, korevaar_targeted_2019}. In the simplest case, $c_\alpha=z_{1-\alpha}$ is the $1-\alpha$ quantile of the standard normal distribution. With this choice, we can define the rejection criterion
\begin{align}\label{eq:cpe_test}
    \varphi_j = 1 
    \ \Longleftrightarrow \ Z_j=\min(Z^{\Se}_j, Z^{\Sp}_j) > c_\alpha
    \ \Longleftrightarrow \ Z^{\Se}_j > c_\alpha \wedge Z^{\Sp}_j > c_\alpha.
\end{align}
This approach gives approximate control of the type I error for $H_{0,j}$ under arbitrary parameter configurations $(\Se_j, \Sp_j)$. Due to asymptotic nature of the procedure, the type I error is often inflated for small sample sizes and proportions close to 0 or 1 \citep{newcombe_two-sided_1998}.
Most importantly, even in the asymptotic case, it does not provide control of the family-wise error rate (FWER) for $H_0$ which is defined as probability to obtain at least one false positive rejection $\varphi_j=1$ of one of the true null hypotheses $H_{0,j}$, $j\in1,\ldots,m$.

Various possibilities exist to allow for control of the FWER of which several are investigated in this work. All methods are based on existing approaches which however need to be adapted to the special structure of the hypothesis problem \eqref{eq:null_global}. The required adaptation is similar for all methods and implies a higher power compared to a native application. This is illustrated in detail based on corresponding comparison regions in Section \ref{sec:regions}. In the following, we briefly describe all investigated methods.

\begin{itemize}
    \item \textbf{no adjustment}: As outlined above, this naive approach is based on \eqref{eq:cpe_test} and $c_\alpha=z_{1-\alpha}$ chosen to be the $1-\alpha$ quantile of the standard normal distribution.
    \item \textbf{Bonferroni}: This well-known multiplicity correction can also be described by \eqref{eq:cpe_test} but $c_\alpha=z_{1-\alpha/m}$ now corresponds to the standard normal quantile for an adjusted local significance level $\alpha^*=\alpha/m$
    \item \textbf{maxT}: This asymptotic parametric approach is based on a multivariate normal approximation of the relevant vector of test statistics \citep{westphal_multiple_2022}.
    The critical value $c_\alpha$ in \eqref{eq:cpe_test} is calculated as an equicoordinate quantile of the multivariate normal distribution with expectation $\mathbf{0}$ and estimated correlation matrix $\mathbf{\widehat{R}}$. Positively correlated diagnostic test decisions imply a smaller $c_\alpha$ and thus less conservative test results.
    \item \textbf{pairs Bootstrap}: A nonparametric approach based on \eqref{eq:cpe_test} whereby $c_\alpha$ is chosen to be the empirical $1-\alpha$ quantile of a bootstrap sample of the maximum test statistic $Z=\max_j(Z_j)=\max_j\min(Z^{\Se}_j,Z^{\Sp}_j)$. For that matter, bootstrap samples of the original data are drawn in a 'paired' fashion such that the original correlation structure of the problem is replicated. 
    A general introduction to this and the next bootstrap approach in the context of regression models is given by \citet{flachaire_bootstrapping_2005}.
    \item \textbf{wild Bootstrap}: This approach is similar to the pairs bootstrap but the mechanism to draw new bootstrap samples is different \citep{flachaire_bootstrapping_2005, zapf_wild_2015}. This approach is investigated although it is not designed for binary data as bootstrap redraws are not necessarily binary.
    \item \textbf{mBeta}: A Bayesian approach whereby two independent multivariate beta-binomial models are fitted for diseased and healthy subgroups \citep{westphal_simultaneous_2019}. A rather particular loss function needs to be employed to achieve (frequentist) control of the FWER relevant for hypothesis problem \eqref{eq:null_global}.
\end{itemize}

All methods are implemented in the new R package \texttt{cases}. Implementation details can be checked in the open source code and corresponding documentation. Besides the Bayesian approach, all methods are capable to provide adjusted p-values for decision making. 

The above-mentioned hypotheses \eqref{eq:null_combined} can be adapted to the case that all index tests $T_j$ are compared to a common comparator $T_0$, i.e. established diagnostic test with unknown diagnostic accuracy, instead of fixed performance thresholds $\se_0$ and $\Sp_0$. All statistical test procedures listed above are also capable to perform more general contrast tests and allow to specify non-inferiority or superiority margins. In the simulation study in Section \ref{sec:numerical}, we however only consider the scenario described in Section \ref{sec:model}.

\subsection{Confidence and comparison regions}\label{sec:regions}

Besides multiplicity adjusted test decisions and p-values, a (two-dimensional) uncertainty quantification is also important in diagnostic accuracy studies. Confidence regions are the multivariate counterpart to confidence intervals. In our case of $m$ index tests, each with unknown sensitivity and specificity, we are interested in $2m$-dimensional confidence regions $\bm C = \bm C_{1-\alpha}$ with coverage probability $1-\alpha$, i.e. we require
\begin{align}\label{eq:ci}
   \forall \bm \theta = (\Sebm, \Spbm) = (\Se_1,\ldots,\Se_m,\Sp_1,\ldots,\Sp_m) \in [0,1]^{2m}:\quad \pr\left( \bm \theta \in \bm C  \right) \geq 1-\alpha.
\end{align}
We focus our attention on rectangular confidence regions $\bm C$ which are the Cartesian product of test specific confidence regions $\bm C_j =  C_j^{\Se} \times C_j^{\Sp}$ with $C_{j}^{\Se} \subset [0,1]$ and $C_{j}^{\Sp} \subset [0,1]$. For simplicity, we only consider one-sided intervals/regions in the following, with all upper limits equal to one.

The duality of confidence intervals and statistical tests for a single parameter is well-known. Similarly, we might expect that defining a statistical test by rejecting $H_j$ when both $\Se_0 \notin C_j^{\Se}$ and $\Sp_0 \notin C_j^{\Sp}$ from \eqref{eq:null_combined} gives us valid, multiplicity adjusted test decisions. While this approach indeed allows (approximate) control of the FWER, we will illustrate in the following that this procedure is conservative and can easily be improved.

\citet{eckert_use_2020} introduce the concept of comparison regions to display the uncertainty in sensitivity and specificity of a diagnostic procedure. 
We define the region of interest $ \bm R = \{(\Se, \Sp) \in [0,1]^2: \Se > \Se_0 \wedge \Sp > \Sp_0 \}$ as the complement of $H_{0,j}$ in $[0,1]^2$. 
A comparison (or decision) region for a single index test is a region $ \bm D_j$ such that defining a statistical test $\varphi_j$ via
\begin{align}
  \varphi_j=1 \ \Longleftrightarrow \ \bm D_j \subset \bm R
\end{align}
allows control of the type I error rate. In other words, a statistical test decision based on the comparison region indicates a significant result if $ \bm D_j$ is completely contained in $\bm R$. In contrast to \citet{eckert_use_2020}, we focus on rectangular regions of interest in this work and in addition introduce the concept of multiplicity adjusted comparison regions. 

The simplest way to illustrate the difference between confidence and comparison regions is the Bonferroni adjustment. For a single index test ($j=m=1$), consider the region
\begin{align}
   \bm A_{j, \alpha^*} = \left( \Sehat_j-c_{\alpha^*} \sqrt{\frac{\Sehat_j(1-\Sehat_j)}{n_1}}  ,1\right] \times 
    \left( \Sphat_j-c_{\alpha^*} \sqrt{\frac{\Sphat_j(1-\Sphat_j)}{n_0}}  ,1\right],
\end{align}
with $c_{\alpha^*}=z_{1-\alpha^*}$ being the $1-\alpha^*$ standard normal quantile. 
Then, $\bm C_{j, \alpha} = \bm A_{j, \alpha/2}$ ($\alpha^*=\alpha/2$) defines an (asymptotic) confidence region for $\bm \theta_j =(\Se_j, \Sp_j)$.
An adjustment for multiplicity is needed here, as both, sensitivity and specificity shall be covered by $\bm C$ with high probability, compare \eqref{eq:ci}.
On the other hand, the comparison region, as the statistical test for the co-primary endpoint problem, does not require an adjustment for $\alpha$ and hence $\bm D_{j, \alpha} = \bm A_{j, \alpha}$ ($\alpha^*=\alpha$).

This directly extends to multiplicity adjusted confidence and comparison regions when $m>1$ index tests are simultaneously evaluated. Here, the Bonferroni adjustment is $\alpha^*=\alpha/(2m)$ for the confidence region, as $2m$ parameters shall be covered.
The adjustment for the corresponding comparison region however is only $\alpha^*=\alpha/m$.
All statistical methods introduced in Section \ref{sec:methods} allow to calculate multiplicity adjusted confidence and comparison regions. For all procedures, one will find that $\bm D_\alpha \subset \bm C_\alpha$ and thus decisions based on confidence regions will not be dual to test decisions but rather more conservative. 

\begin{figure}
	\centering
	\includegraphics[width=0.48\linewidth]{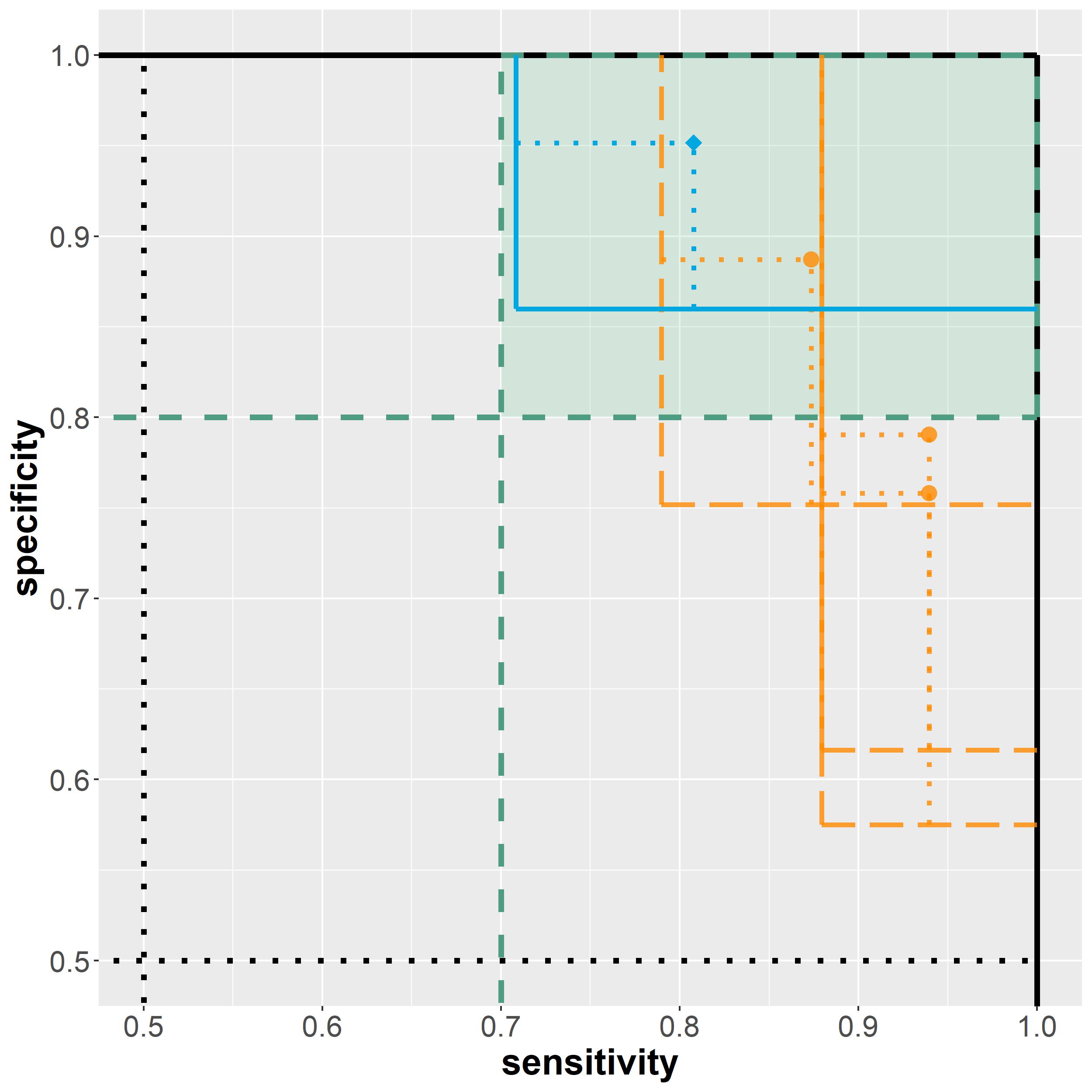}
	\includegraphics[width=0.48\linewidth]{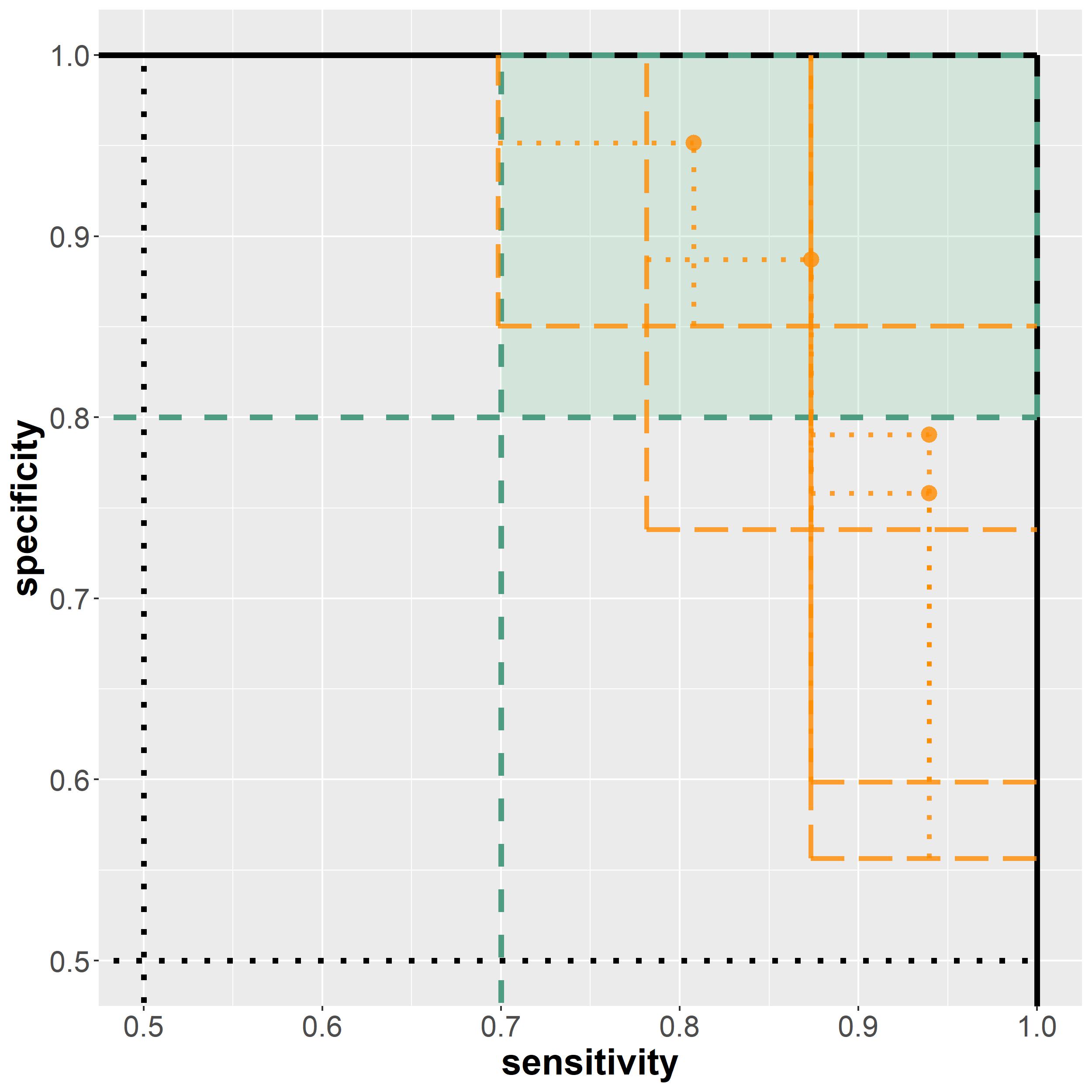}
	\caption{Exemplary analysis of synthetic example with four index tests. Region of interest ($\Se_0=0.8$, $\Sp_0=0.7$) is displayed in green. 
		Left: comparison regions. Right: confidence regions. Solid/blue lines imply a rejected null hypothesis; dashed/orange lines imply a non-rejected null hypothesis.}
	\label{fig:regions}
\end{figure}

Figure \ref{fig:regions} illustrates this difference for a synthetic data set with $n_1=30$ cases and $n_0=90$ controls and $m=4$ index tests with varying accuracy. The region of interest $\bm R$ is displayed in the upper right-hand corner in green. In the left subfigure, multiplicity adjusted comparison regions by the maxT-approach are indicated by solid lines around each point estimate $(\Sehat_j, \Sphat_j)$. The study conclusion in this case would be a rejection of the null $H_{0,j}$ only one of the four index tests ($j=1$, solid/blue lines) as $\bm D_1 \subset \bm R$ but $\bm D_j \not \subset \bm R$ for all other tests $j=2,3,4$ (dashed/orange lines). The right subfigure reveals that a decision based on the corresponding (maxT-adjusted) confidence regions would have not resulted in any rejection of the null hypothesis as $\bm C_j \not\subset \bm R$ for all $j=1,2,3,4$.

\subsection{Simulation study}\label{sec:numerical}

We perform a simulation study to evaluate and compare the multiple comparison procedures described in Section \ref{sec:tests}. Our primary goal is a comparison with regards to family-wise error rate and statistical power. The (disjunctive) statistical power is the probability to obtain at least one rejection of a false null hypothesis $H_{0,j}$. For that matter, two different data-generating mechanisms ('LFC', 'Biomarker') are investigated and outlined in the following. The key features of the simulation study are described in Table \ref{tab:sim_features} following the ADEMP framework \citep{morris_using_2019, boulesteix_introduction_2020}.

\begin{table}[]
    \begin{tabular}{p{0.15\textwidth} p{0.75\textwidth}}
        \toprule
        \textbf{Feature} &  \textbf{Description}\\
        \midrule 
        \rowcolor{black!10}
        Aims &  Comparison of different multiple comparison procedures (MCPs)\\
        Data generating mechanisms & Different scenarios (w.r.t. sample size, control-to-case ratio, correlation structure, ...) in two distinct settings ('LFC': Section \ref{sec:sim_lfc}, 'Biomarker': Section \ref{sec:sim_bio})  \\
        \rowcolor{black!10}
        Methods of \newline analysis &  Multiple comparison procedures for hypothesis problem \eqref{eq:null_global}, described in Section \ref{sec:tests}\\
        Performance measures & Family-wise error rate (FWER) and (disjunctive) power\\
        \rowcolor{black!10}
        Number of \newline repetitions &  $n_{sim}=10,000$ per scenario\\
        \bottomrule
    \end{tabular}
    \caption{Key features of simulation study}
    \label{tab:sim_features}
\end{table}

For both simulation settings (compare Section \ref{sec:sim_lfc} and \ref{sec:sim_bio}), many different scenarios are considered which are described by a number of design parameters (setting, sample size, prevalence, correlation structure, ...). For each scenario, $n_{sim}=10,000$ synthetic datasets were generated. The simulation study is conducted such that all methods are applied to the exact same synthetic datasets. 
The standard (Monte-Carlo) error for estimated proportions $p$ such as the FWER or statistical power is $(p(1-p)/n_{sim})^{1/2}$ which is bounded by $0.005$ for $n_{sim} = 10,000$.
In all simulation scenarios, we investigated sampling with a fixed sample sizes $n_0$, $n_1$ which corresponds to a simple case-control within-subjects study design.

All numerical experiments presented in this work are reproducible by means of the new R package \texttt{cases}. Instruction for that matter are provided via an additional simulation repository which also contains the results of additional sensitivity analyses\footnote{\url{https://maxwestphal.github.io/cases_simstudy/}, last accessed \today.}. 
As the source code for all statistical methods is available in the \texttt{cases} package, this also serves as a documentation of implementation-related details. 
To reduce the computational demand of the simulation study, we consider only the comparison of $m$ candidate tests to fixed performance criteria $\Se_0 = \Sp_0$. The \texttt{cases} package furthermore supports comparison of $m$ candidate tests to the unknown sensitivity and specificity of a comparator and other relevant contrast tests. 
The simulations were conducted in R (version 4.2.1) with help of the \texttt{batchtools}\footnote{\url{https://CRAN.R-project.org/package=batchtools}, last accessed \today.} package.

\subsubsection{LFC setting}\label{sec:sim_lfc}

In the first setting, a single synthetic data set consists of two separate samples for diseased and healthy. For that matter, two different multivariate Binomial models are employed to generate binary data matrices $\bm Q^{1}$ and $\bm Q^{0}$. These matrices have entries
\begin{align}
    Q_{ij}^g= \one(T_{ij} = D_i),\quad i=1,\ldots,n,\quad j=1,\ldots,m, \quad g=0,1,
\end{align}
whereby $Q_{ij}^g$ indicates if the $j$-th diagnostic test (data-based decision rule) $T_j$ under consideration correctly predicted the $i$-th disease status $D_i$ in subgroup $g\in \{0,1\}$. The multivariate binomial distribution in general has $2^m$ parameters but reduced representations do exist which can be specified by the mean vector and correlation matrix. We utilize the \texttt{R} package \texttt{bindata}\footnote{\url{https://CRAN.R-project.org/package=bindata}, last accessed \today.} to generate the data $Q^{\Se}$, $Q^{\Sp}$ with given first moments $\Sebm$, $\Spbm$, respectively \citep{leisch_generation_1998}.

In this scenario, we focus on least-favorable parameter configurations (LFC) which are defined as parameters which maximize the FWER. As discussed by \citet{westphal_evaluation_2020}, an LFC in this case is defined by mean vectors $\Sebm$, $\Spbm$ with entries

\begin{align}
    \Se_j = \begin{cases} \Se_0, &b_j=1 \\ 1, &b_j=0 \end{cases}, \quad \Sp_j = \begin{cases} \Sp_0, &b_j=0 \\ 1, &b_j=1 \end{cases},
\end{align}

whereby $\Se_0$, $\Sp_0$ are the parameter thresholds which define the hypotheses \eqref{eq:null} and $\bb=(b_1,\ldots,b_m) \in \{0,1\}^m$ is a fixed binary vector. For each index $j$ we thus have either $\Se_j$ or $\Sp_j$ exactly at the boundary of the null hypothesis and the other parameter equal to one. For each of the two endpoints, a correlation matrix is specified in addition. To limit computational resources, we restricted the attention to equicorrelation between non-degenerated parameters, i.e. $\rho^{se}_{j j'} = \rho^{\Se}$ for all indices $j \neq j'$ with $\Se_j \neq 1$ and $\Se_{j'} \neq 1$.

\subsubsection{Biomarker setting}\label{sec:sim_bio}

The least-favorable parameter (LFC) setting discussed in the last section allows to assess worst case error rates. However, LFCs are rarely representative of real-world situations \citep{westphal_multiple_2022}. The goal of this second simulation study is thus to evaluate the multiple comparison procedures under more realistic parameter configurations. 

To generate a single synthetic data set, we start out by simulating $l$ continuous biomarkers from a multivariate binormal ROC model \cite[Section\,4.4]{pepe_statistical_2003}. 
This means, that we consider multivariate markers $\bm V= (V_1,\ldots,V_l)$ such that
\begin{align}\label{eq:score_def}
    \bm V \given D=1 \sim \setN(\bm \mu^1, \bm R^1), \quad  \bm V\given D=0 \sim \setN(\bm 0, \bm R^0).
\end{align}
Hereby, $\mu^1_k$, is the mean of marker $k$ in the diseased ($D=1$) population. For simplicity, we assume a mean $\mu^0_k=0$ for all $k=1,\ldots,l$ in the healthy population ($D=0$) and a variance of one in both populations. If we aim for a certain AUC of marker $k$, this can be achieved by solving
\begin{align}
    \AUC_k = \Phi \left( \frac{\mu_k^1 }{ \sqrt{2}} \right)
\end{align}
for $\mu^1_k$ which has the solution $\mu_k^{1*}=\sqrt{2}\Phi^{-1}(\AUC_k)$ \cite[Section\,4.4.2]{pepe_statistical_2003}. Hereby, $\Phi$ is the cumulative distribution function of the standard normal distribution. 

These continuous scores $V_k$ are split at different cut points $c_{j}$ to obtain an overall list of diagnostic tests $T_j = \one(V_{k_j} > c_j)$, $j=1,\ldots,m$ with $m \geq l$. The resulting decision rules have true parameters
\begin{align}
    \Se_j = \Phi \left( \mu^1_{k_j}-c_j \right), \quad \Sp_j = 1- \Phi \left( -c_j \right)
\end{align}
depending on the cut point $c_j$. Alternatively, this process can also be identified as thresholding risk model scores to obtain different prediction models. 
The correlation matrices $\bm R^0, \bm R^1$ in \eqref{eq:score_def} have a simple structure (e.g. equicorrelation) in our simulation study. Together with the chosen cut points, this induces a certain correlation structure for the actual binary decision rules. Naturally, two (close) nearby cut points, induce (highly) similar predictions and thus a (highly) positive correlation between estimates.

\section{Results}\label{sec:results}

\subsection{Simulation study}\label{sec:results_sim}

\subsubsection{LFC setting}

\begin{figure}[t!]
	\centering
	\includegraphics[width = 0.48\linewidth]{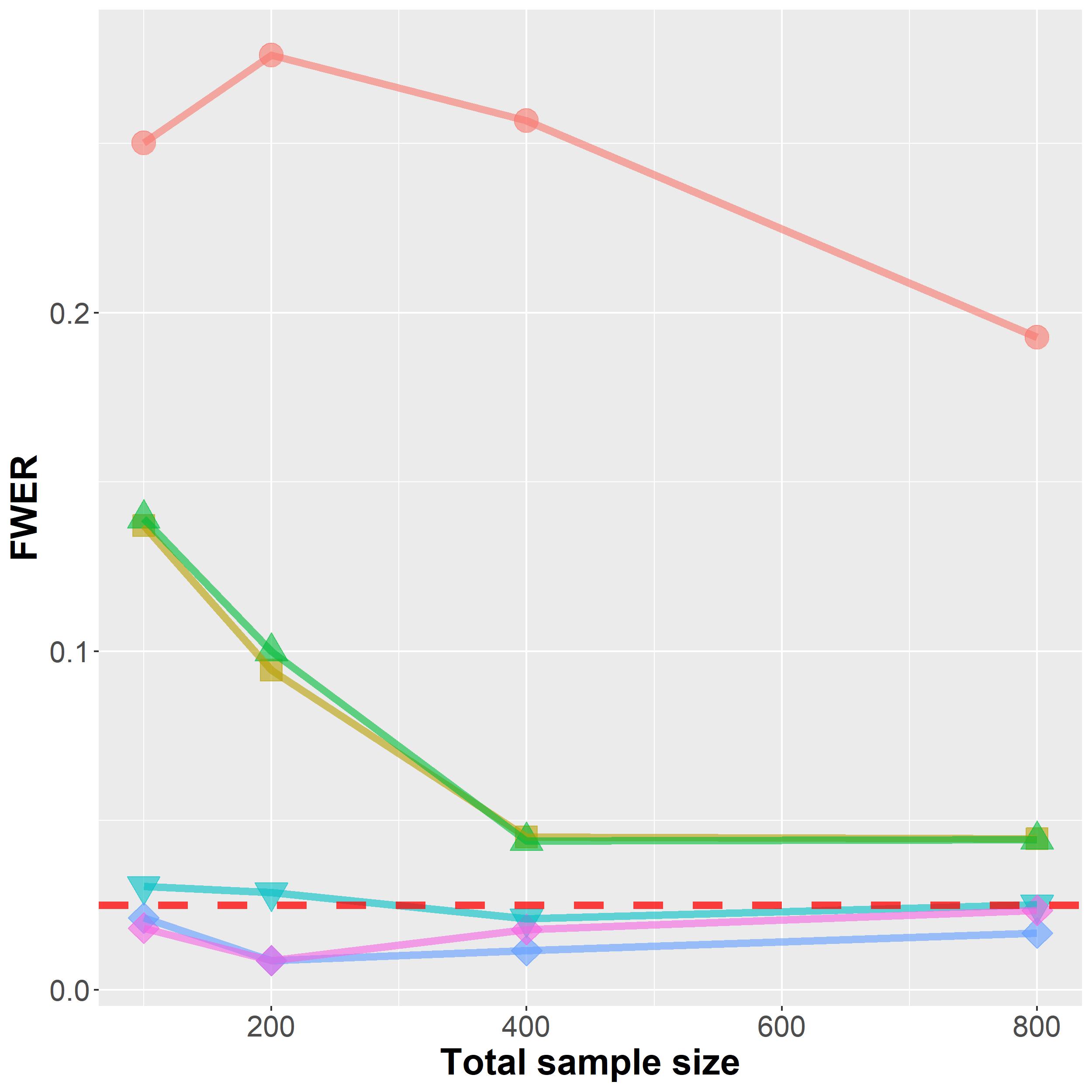}
	\includegraphics[width = 0.48 \linewidth]{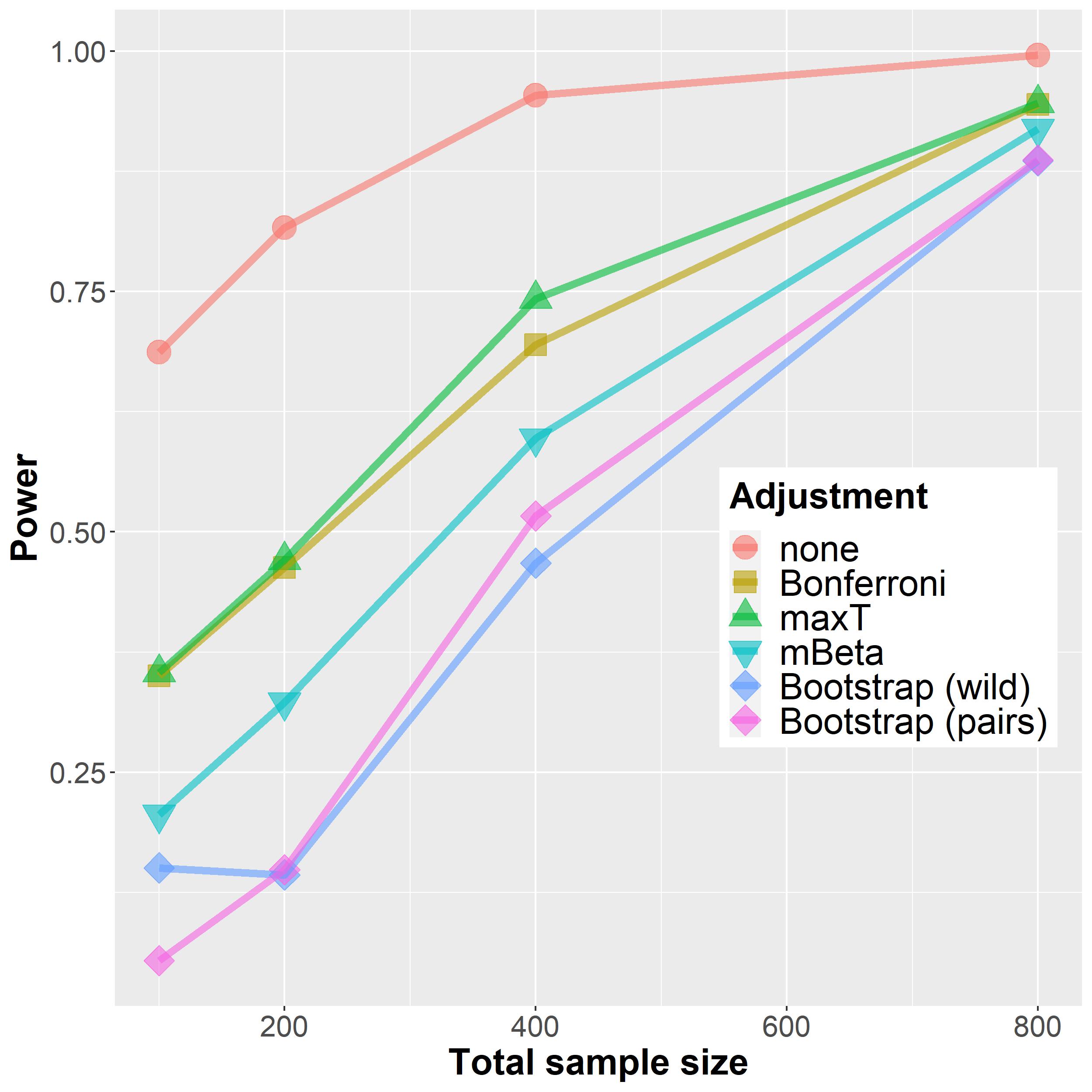}
	\caption{Simulation results for the 'LFC' setting. Left: FWER, Right: Power. The horizontal axis shows the total sample size (cases and controls $n=n_1+n_0$) in both subplots.}
	\label{fig:sim_results_lfc}
\end{figure}

The results from the LFC (least-favorable configuration) setting described in Section \ref{sec:sim_lfc} are displayed in Figure \ref{fig:sim_results_lfc}. On the left, the FWER is shown depending on the total sample size $n=n_1+n_0$ for a control-to-case ratio of $3:1$ and $m=10$ index tests. The parameter values are set to an LFC with $\Se_0=\Sp_0=0.8$, compare Section \ref{sec:sim_lfc}. The red dashed vertical line shows the (one-sided) target significance level of $\alpha=0.025$.

First, applying no multiplicity adjustment at all, clearly leads to a vastly increased FWER of $20\%$ and more compared to the target significance level of $\alpha=0.025$. Second, the asymptotic methods (maxT, Bonferroni) are capable of controlling the FWER for large $n$ but fail to do so for small $n$. For $n=100$ both methods show an increased FWER of up to 10\%, declining to around $5\%$ for $n \geq 800$. Third, the Boostrap approaches (pairs, wild) and the Bayesian mBeta approach all control the FWER for all but the lowest sample size of $n=100$.

The right part of Figure \ref{fig:sim_results_lfc} shows the (disjunctive) power for the same data instances after dropping the minimal acceptance criteria $\Se_0, \Sp_0$ both to $0.75$. The ordering of methods stays the same as for the FWER analysis, as expected. The gap between mBeta and Bootstrap approaches is somewhat more pronounced than before.

Further analyses (not shown visually here) indicate, that these findings for FWER and power remain qualitatively similar when only $m=5$ index tests are considered are when the correlation structure between diagnostic test result is changed. When the control-to-cases ratio is changed to $10:1$, all curves are essentially shifted. To obtain a similar power as before, a higher total sample size is required due to the dependence on $\min(n_1,n_0)$. As expected, when $\Se_0$ is changed from 0.8 to 0.9, the FWER is increased for finite samples but the qualitative observations regarding the asymptotic behaviour still persist.

\begin{figure}[t!]
	\centering
	\includegraphics[width = 0.48\linewidth]{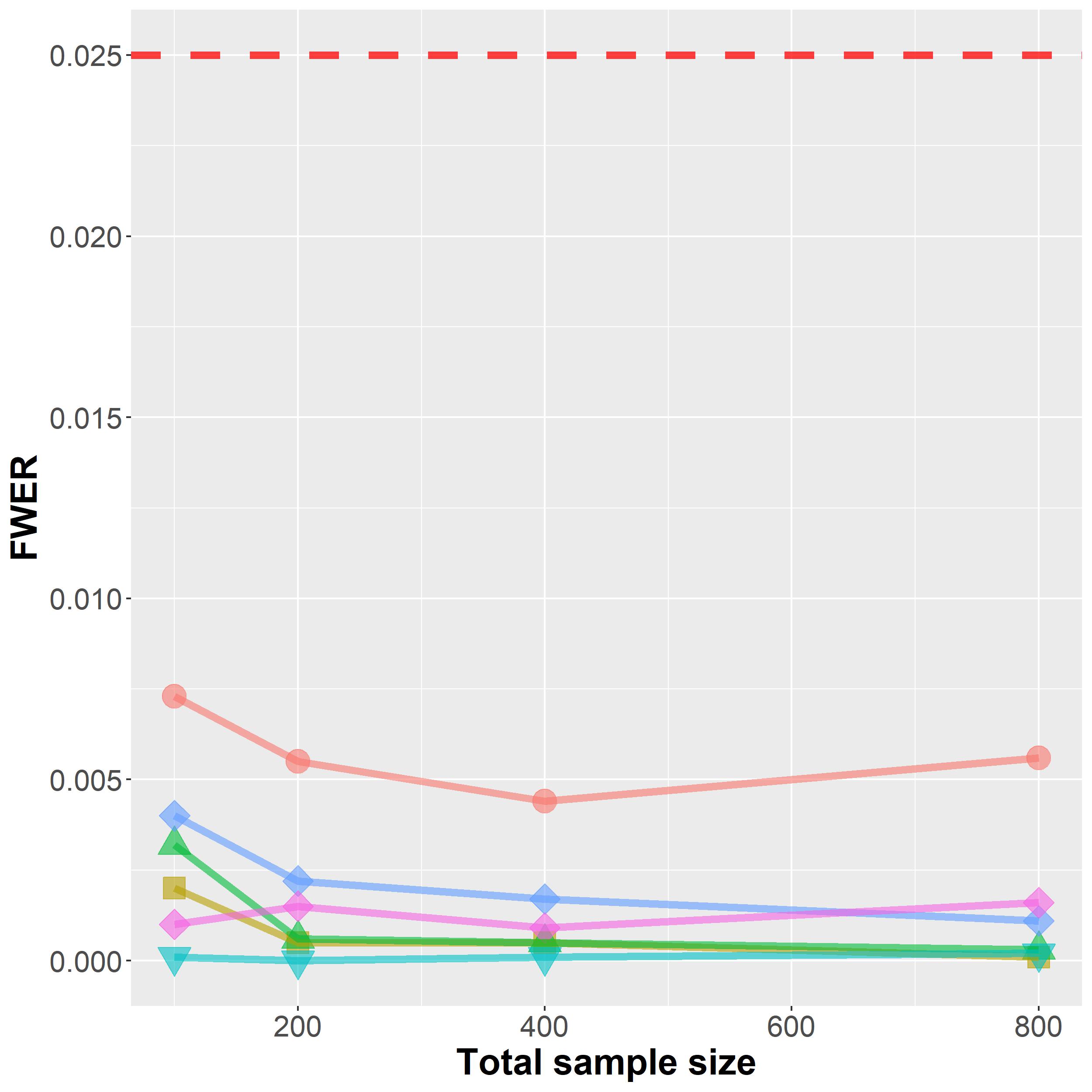}
	\includegraphics[width = 0.48 \linewidth]{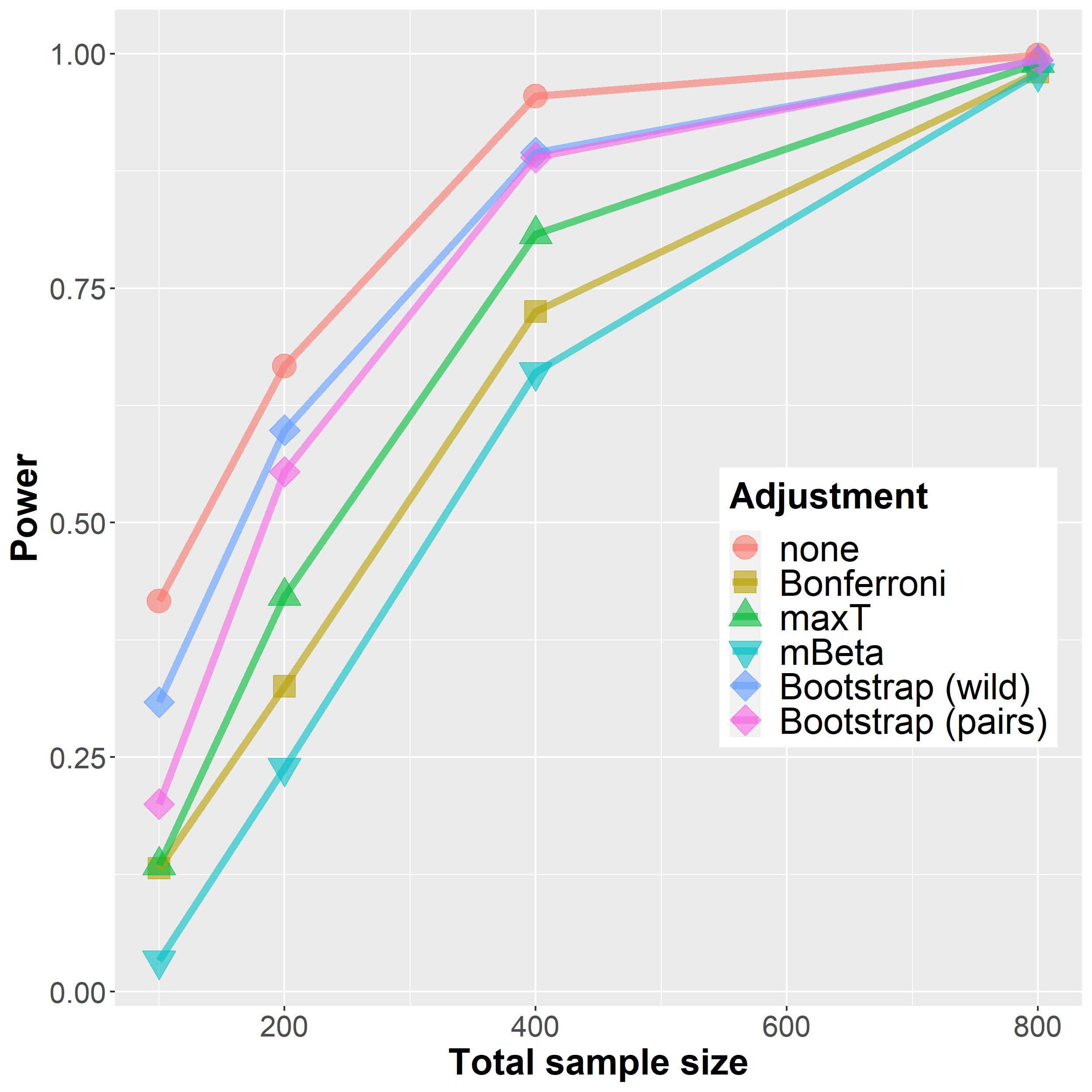}
	\caption{Simulation results for the 'Biomarker' setting. Left: FWER, Right: Power. The horizontal axis shows the total sample size (cases and controls $n=n_1+n_0$) in both subplots.}
	\label{fig:sim_results_bio}
\end{figure}

\subsubsection{Biomarker setting}

Figure \ref{fig:sim_results_bio} shows results from the simulation in the more realistic biomarker setting described in Section \ref{sec:sim_bio}. In contrast to the worst case assessment in the last section, the FWER is below $0.005$ for all methods and nearly all sample sizes and as such far below the target significance level of $\alpha=0.025$. This behaviour is expected based on previous work \citep{westphal_evaluation_2020}.

Regarding statistical power the ordering of methods is mostly similar to the situation in the LFC setting. One important difference is the much better performance of both Bootstrap approaches which seem to adapt much better to the underlying generative distribution in this situation.

\subsection{Analysis of motivating example}\label{sec:results_rwd}

To conclude, we apply our methodology to real-world data and interpret the results. For that matter, the pairs bootstrap approach is applied in the two motivating example scenarios introduced in Section \ref{sec:example}. To formalize our requirement that a high sensitivity is prioritized, we define the minimal acceptance criteria as
$\Se_0=0.9$ and $\Sp_0=0.7$. The entire analysis can be reproduced by following the corresponding vignette of the new R package \texttt{cases}.

\subsubsection{Scenario A: Biomarker assessment}

The most promising of the 15 investigated classification rules in terms of diagnostic accuracy on the evaluation data (212 cases, 357 controls) is the maximal area of any present cell nucleus with a threshold of $700 \mu m^2$ applied for categorization into cases ($> 700 \mu m^2$) and controls ($\leq 700 \mu m^2$). This rule achieves an empirical sensitivity of $96.0\%$ and an empirical specificity
$80.6\%$. The multiplicity adjusted lower comparison bounds are $92.1\%$ and $74.6\%$, respectively. As both lower bounds are larger than their respective accepetance criteria, the co-primary null hypothesis can be rejected for this test candidate. This is visualized in Figure \ref{fig:example} (left). None of the other comparison regions is completely contained in the region of interest corresponding to the alternative hypothesis. In effect, the null hypothesis cannot be rejected for any other test candidate even though individual sensitivity and specificity estimates of other candidates are larger than those reported above.

\subsubsection{Scenario B: Risk model evaluation}

\begin{figure}
	\centering
	\includegraphics[width=0.48\linewidth]{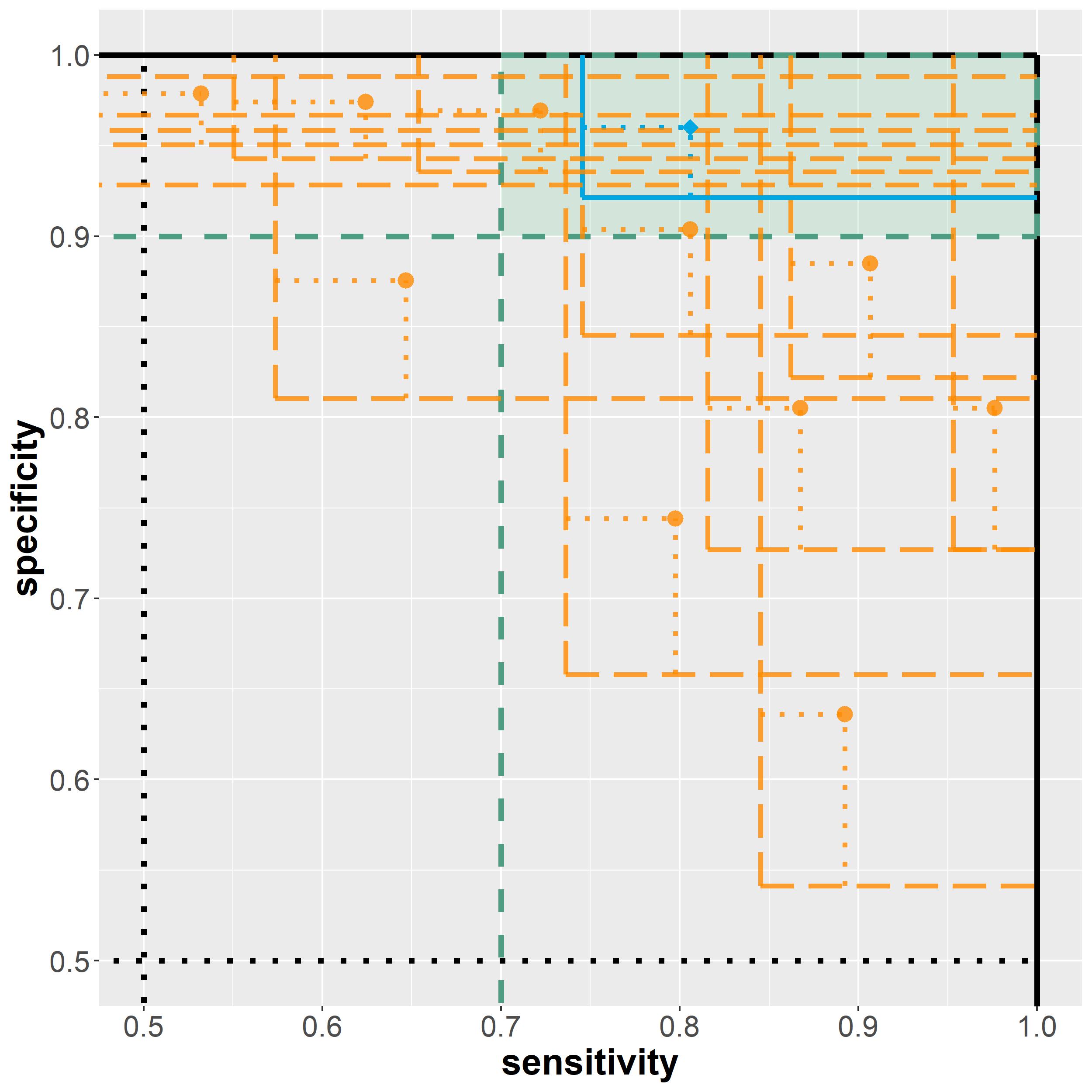}
	\includegraphics[width=0.48\linewidth]{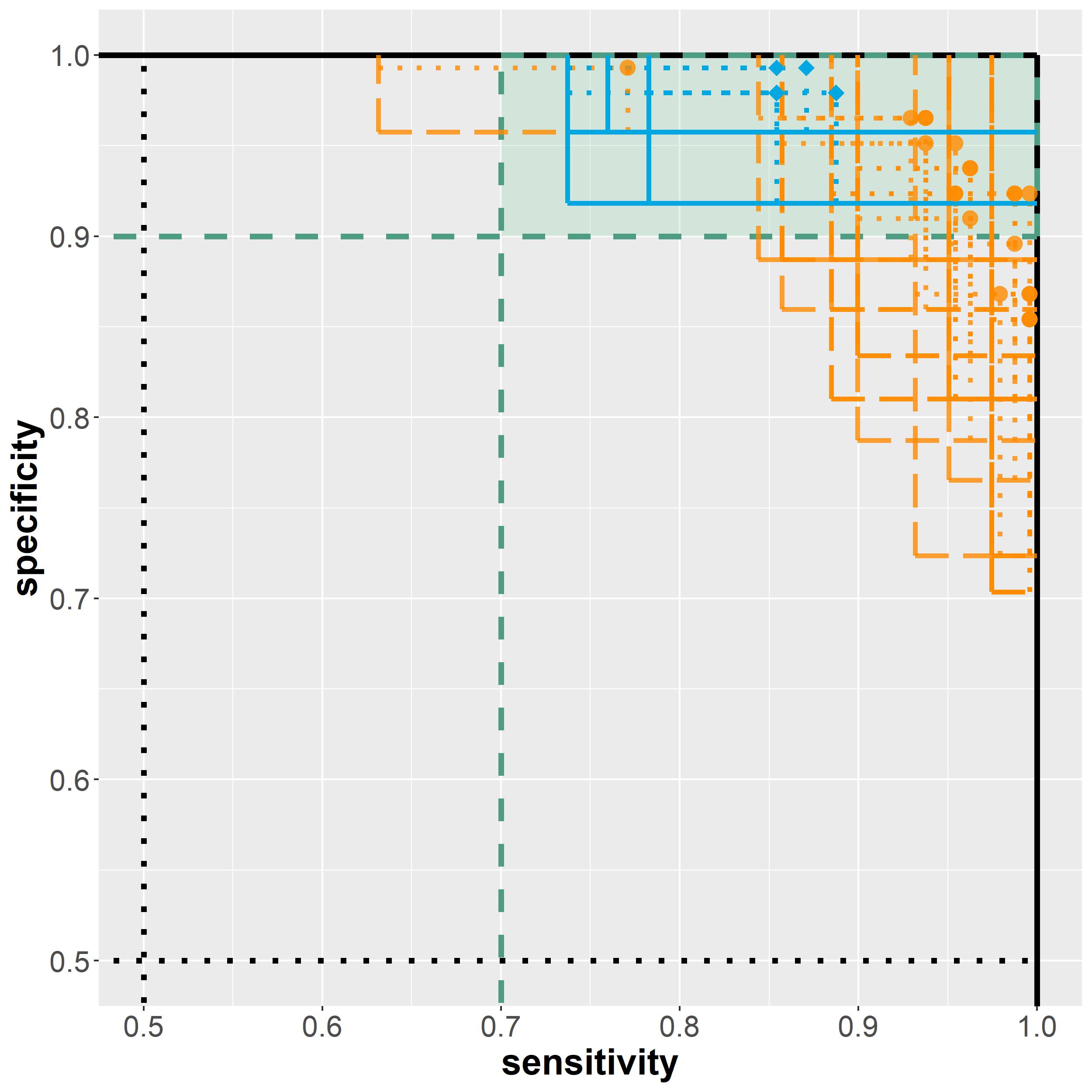}
	\caption{Results from the analysis of  the two breast cancer diagnosis example scenarios. Left: biomarker assessment (scenario A). Right: risk model evaluation (scenario B). Solid/blue lines imply a rejected null hypothesis; dashed/orange lines imply a non-rejected null hypothesis.}
	\label{fig:example}
\end{figure}

In total, 25 risk prediction models were assessed on the evaluation data (71 cases, 119 controls). For four models, the null hypothesis $\Se \leq 0.9 \vee \Sp \leq 0.7$ can be rejected as these models have lower comparison bounds for sensitivity and specificity that greater than the corresponding acceptance criteria.
Empirical sensitivities of these models lie between $99.3\%$ and $97.9\%$ and specificities between $88.8\%$ and $85.4\%$. This is visualized in Figure \ref{fig:example} (right).
These four classifiers all result from thresholding different model risk scores at the lowest probability threshold of $0.10$. Because all four models are deemed acceptable regarding their discriminatory performance, other criteria may influence the final model decision. For instance, the most parsimonious model can be chosen to increase interpretability. In this example, the lasso fit (elastic net mixing parameter $\alpha=1$) results in only 11 non-zero regression coefficients and is much sparser as the three competitors (with $\alpha < 1$) which have 18, 19 and 23 non-zero coefficients, respectively.

\section{Discussion}\label{sec:discussion}

In this work, we investigated statistical inference methods for diagnostic accuracy studies with multiple index tests and co-primary endpoints sensitivity and specificity. Multiplicity corrections are relevant in this context as omitting a suitable correction can otherwise induce overoptimistic results and inflated error rates. While a control is easily possible using a traditional (FWER) correction for all $2m$ parameters ($m$ index tests with unknown sensitivity and specificity), this approach is too conservative for the co-primary endpoint analysis. Several multiple comparison procedures can be adapted to the specific testing problem such that this conservative nature is resolved. This difference was illustrated by contrasting confidence regions and comparison regions in Section \ref{sec:regions}. Only the latter allow for a two-dimensional uncertainty quantification which is dual to the statistical test result.

We conducted an extensive simulation study to compare five multiple comparison procedures with regards to family-wise error rate (FWER) and statistical power. All five procedure are capable to control the FWER at least asymptotically. The Bonferroni adjustment and maxT-approach both need quite large sample sizes to reach the target significance level. This however depends crucially on the control-to-cases ratio. A more skewed class distribution results in higher sample size demand. The Bayesian mBeta approach and the two Bootstrap (pairs, wild) procedures managed to control the FWER much better under least-favorable parameter configurations. Based on the simulation study, we recommend the use of the pairs Bootstrap approach. This approach yielded good FWER control for small sample sizes and competitive power in realistic scenarios. The wild bootstrap approach has very similar performance, but the method is more complex, has more design choices, and was not originally designed to be used for binary data. For larger sample sizes, the maxT-approach is a competitive alternative and easier to apply compared to the bootstrap procedures. The Bonferroni method is by far the easiest method to apply but too conservative in some situations and as such suboptimal. 

In this work, we focused on the family-wise error rate and the statistical power. In our framework, point estimates are slightly shrunk due to the implemented minor regularization to avoid singular (co)variance estimates. However, point estimates were not adjusted for multiplicity. \citet{westphal_multiple_2022, westphal_evaluation_2020} illustrated how to apply the maxT-approach to obtain median-conservative point estimators. This generic approach can be also adapted to the other multiple comparison procedures. 

All investigated methods can be adapted to the case that not only two (diseased and healthy) subpopulations need to be distinguished but more than two subpopulations, e.g. different disease severities. This capability is already implemented in the \texttt{cases} package. We expect that the performance will then crucially depend on the prevalence of the smallest subpopulation. This should however be confirmed in dedicated simulation studies.

\section*{Conflict of interest}
The authors declare that there is no conflict of interest.

\bibliography{literature}

\end{document}